# Photodetachment Microscope with Repulsive Coulomb Field

P.A. Golovinski and A.A. Drobyshev

*Physics Research Laboratory, VSUAC, ul. 20-letiya Oktyabrya 84, Voronezh, 394006 Russia*
e-mail: golovinski@bk.ru

Investigation of electronic waves with high coherence in photodetachment of a negative ion gives a physical basis to develop the holographic electronic microscopy with high resolution. The interference pattern is considered in the framework of steady-state wave approach. In semiclassical approximation, an outgoing wave is described by the amplitude slowly varying along a trajectory. Quantum description of electron photodetachment from negative ion is formulated with the help of the inhomogeneous Schrödinger equation. Its asymptotic solution is expressed in terms of the Green function that has exact expression for the homogeneous electric field and the Coulomb field. It is demonstrated that repulsive Coulomb field is effective for magnification of the interference pattern at a short distance from an ion. For the first time, as shown for this case, the interference pattern in asymptotic area can be calculated by means of global semiclassical approximation or, a little more roughly, by simple uniform field approximation. New proposals and possible parameters for experimental measurement scheme for the effect are discussed.

PACS number(s): 32.80.Gc, 42.40.Kw

## 1. INTRODUCTION

In recent years an increasing attention is focused on the physical effects of quantum interference. The principle problems of quantum measurements and some perspective applications became the basic stimulus to investigate this area of researches [1, 2]. The electronic waves with high degree of coherence became a basis for progress in electronic microscopy with high resolution, and for the developing of the holography with a low energy of electrons. This circumstance opens an opportunity for nondestructive observation with the help of a new tool for nanometer scale objects.

Earlier, the analytical theory was developed to explain experimental results for photodetachment microscope where electrons were detached from negative ions in a homogeneous electric field [3-6]. The advantage of this theory is exact and compact formulation of the result. The quantum interference is taken into account in a three-dimensional space, including a caustic surface and shadow zone. The idea of the photodetachment microscope was proposed in [7, 8]. The principle of this type of microscope is based on two fundamental effects. The first effect is the Wigner law for the electron photodetachment cross section of a negative ion in the vicinity of a threshold. According to this law, the partial cross section for electron photodetachment into a final state with orbital-angular momentum quantum number $l$ is exponential function of electron energy $\varepsilon$ of photoelectron:

$$\sigma \sim \alpha \varepsilon^{l+1/2}. \qquad (1)$$

For negative halogen ion the initial electron state has orbital quantum number $l=1$, and the possible angular moments of the final states are equal 0 or 2. The corresponding cross sections behave as well as $\varepsilon^{1/2}$ and $\varepsilon^{5/2}$. The partial channel of photodetachment into final *s*-state dominates close to the threshold, and this state is described by a spherical outgoing wave. Multiphoton detachment near the threshold may be a source of coherent electrons also [9]. The second effect is an interference of coherent electronic waves at the propagation of monoenergetic electronic beam in an electric field along different paths. For example, in a homogeneous electric field two different trajectories comes from the initial point to the same final point. This dynamics is similar to the particle motion, starting under the angle to the horizon in the homogeneous gravitation field. For the case, when the initial altitude is non zero, and for a given momentum, we have two different ways to hit the same final target in accordance with two different elevation angles. The time of classical flight for these two trajectories are different, but for propagation of steady-state waves the interference conditions are well defined.

The problem of experimental observation of this phenomenon has been solved by the group at the Laboratoire Aimé-Cotton [10-14], by the overlapping the ion beam and the light, so that two separate sets of interference patterns can be detected. Measuring the distance between these patterns it is possible to calculate and extract the Doppler free electron affinity. The photoelectron energy must be very small (in the range 0.01 - 0.4 meV), and the technique is effectively limited to negative atomic ions for which the detachment cross section is high even close to the detachment threshold, namely s-wave detaching ions. Photodetachment microscopy with application to the measurement of electron affinities of neutral atoms and molecules with interference pattern was used in [15-19].

The combination of fields, for another case of electron wave propagation in photoionization of H atom from n=2 state in uniform electric field, has been investigated by semiclassical method and quantum mechanical approach [20, 21]. Nearly the same phenomena, but for xenon atom, was observed experimentally in [22] with only qualitative semiclassical description. The first experimental results of a technique, called photoionization microscopy, were presented. Photoelectrons ejected in threshold photoionization of Xe were detected in a velocity map imaging apparatus, and interference between various trajectories by which the electron moves from the atom to the detector was observed. The main observed features were interpreted within the framework of the semiclassical approximation. The comparison of the experimental and theoretical radial probability distributions demonstrated a large discrepancy. The treatment was strictly limited to the semiclassical approximation, without any correction to the phase and with no account for the failure of the semiclassical approximation when the momentum was vanished. In addition, the contribution of tunneling integrals has been neglected. As a result, the fitting of experimental data was not good enough.

The idea of new configuration with repulsive Coulomb field has been suggested in [23]. It has been shown that such a field can be used for essential magnification of the electron current spot at the detector and the analytical solution with the initial s-wave has been obtained. The purpose of this paper is a more complete description of the electron wave interference after the process of a negative ion photodetachment in the static repulsive Coulomb field. In this context, the process in homogeneous electric field can be interpreted as a limiting case and a global asymptotic approximation at the same time. The parameters of laser and electric field are estimated which are suitable for experimental observation of the effect. The range of validity for assumed approximations is clearly determined.

## 2. SEMICLASSICAL THEORY

In further consideration we assume that previous results of photodetachment cross section calculations in the presence of external static electric field are known [24-28], and also a total yield of electrons. All further attention is concentrated on the interference of electronic waves and photocurrent distribution in space. In particular, the s-wave photodetachment from negative S ions has been studied in external electric fields up to 220 V/cm for laser polarization parallel to the electric field direction [29]. The ratio of the electric field-on to electric field-off cross sections was found to be in excellent agreement with theoretical predictions. As predicted, electron wave-function rescattering effects were very small in both the amplitude and the phase of the electric-field-induced oscillations. For the case of photodetachment from $S^{--}$, s-wave detachment was by far the dominant channel near threshold, and the cross section in zero electric field was given by the Wigner law. The calculations based on [30] indicate that rescattering causes a relative change of the ratio of at most 1.6% at zero energy, and that the deviation rapidly decreases to less than 1% as the electron energy increases above 2 cm$^{-1}$. This relative deviation is far too small to take into account.

The distance $r$ in the vicinity of negative ion where electron motion is approximately free of external electric field action is determined by inequality $1/p \ll r \ll p^2/2F$, where $p$ is electron momentum in the moment of detachment, and $F$ is electric field strength. Atomic units are used unless otherwise noted. Physical meaning of the formulated condition is the work of electric field at the electron wave length distance is less than its kinetic energy. This excludes electron energies too close to the threshold. Thus the overall picture of the interference phenomena can be described as a two step process in the semiclassical approximation [31-33]. In this approach the first step is photodetachment from negative ion into final s-state of electron, with neglecting of the external static field. The second stage is wave propagation that is described with the help of amplitude, slowly varying along a trajectory. The waves going by different paths demonstrate interference. The total electron flux at a fixed point of observation is proportional to the square of partial wave amplitudes along different trajectories. The result can be formulated mathematically in the following way [34]. The Schrödinger equation for a particle in electric field with a potential $U(q)$ is of the form

$$i\frac{\partial \psi(q,t)}{\partial t} = -\frac{\Delta}{2}\psi(q,t) + U(q)\psi(q,t), \qquad (2)$$

where $\psi$ is a wave function. We assume the initial condition for the wave function is of the form

$$\psi\big|_{t=0} = \varphi(q)e^{iS(q)}, \qquad (3)$$

and $S(q)$ is a classical action. Smoothly varying function $\varphi(q)$ is differed from zero only inside a bounded space. Here the classical Hamilton equations are

$$\dot{q} = \frac{\partial H}{\partial p}, \quad \dot{p} = -\frac{\partial H}{\partial q}, \qquad (4)$$

where the Hamilton function is

$$H = \frac{p^2}{2} + U(q). \qquad (5)$$

The initial point in the phase space is $(p_i, q_j)$, and the action $S_j$ along a trajectory, started at the point $(p_i, q_j)$, is

$$S(Q,t) = S(q_j) + \int_0^t L\,d\tau, \quad L = \frac{\dot{q}^2}{2} - U(q), \qquad (6)$$

where $L$ is the Lagrange function.

The solution to Eq. (2) with the initial condition of Eq. (3) has the asymptotic form

$$\psi(Q,t) = \sum_j \varphi(q_j) \left| \frac{DQ}{Dq_j} \right|^{-1/2} e^{iS_j(Q,t) - i\pi\mu_j/2}, \qquad (7)$$

where $DQ/Dq_j$ is a transformation Jacobian for coordinates along a trajectory. Parameter $\mu_j$ is an integer, named the Maslov index of the $j$-th trajectory [35]. At a unary contact of a trajectory with caustic surface the phase changes is $\pi/2$, and the Maslov index is equal 1, without a contact with caustic the index $\mu_j = 0$. This situation is typical for homogeneous field, and for the Coulomb potential. The factor $DQ/Dq_j \sim j_{cl}$, where $j_{cl}$ is a classical density of a current.

The quantum formula for electron density of a current gives

$$\mathbf{j} = \frac{i}{2}(\psi \nabla \psi^* - \psi^* \nabla \psi) \approx |\psi|^2 \nabla S. \qquad (8)$$

We take the phase dependence of a wave function in the semiclassical approximation as a classical action $S$. It follows from the Eq. (8) that for two different trajectories, coming to the same point, the current can be written down in the form

$$j \sim j_{cl} \left| e^{iS_1 - \frac{i\pi}{2}\mu_1} + e^{iS_2 - \frac{i\pi}{2}\mu_2} \right|^2. \qquad (9)$$

If classical current $j_{cl}$ is known, then, to obtain a correct asymptotic formula for the current density $j$, it is essential to determine $S_1$, $S_2$, $\mu_1$, $\mu_2$. For the first trajectory without a contact with a caustic surface the parameter $\mu_1 = 0$, and for the second trajectory the index $\mu_2 = 1$. As a result, we have

$$j \sim j_{cl}(1 + \cos\alpha), \qquad (10)$$

where $\alpha = S_1 - S_2 + \pi/2$. Eq. (10) describes the interference pattern with a number of bright and dark rings. However, at the caustic surface a non physical singularity is predicted because of a contact of two trajectories, starting from the same point with a fixed energy of electrons and with different initial directions.

### 3. QUANTUM-MECHANICAL APPROACH

The complete experimental picture can be reproduced by the advanced quantum theory. We describe the photodetachment of electron from a negative ion by a weak monochromatic laser field with linear polarization. The influence of laser field can be assumed small and is considered by the perturbation theory. Therefore, the photodetachment of electron is a steady-state process. An initial state of a negative ion is only slightly modified by photocurrent. The electron interaction with a laser field can be written down in the dipole approximation as [36]

$$W = -\frac{iA}{c}(\mathbf{e}\nabla), \qquad (11)$$

where $A$ is an amplitude of the vector potential of a laser field, $c$ is a speed of light in vacuum, $\mathbf{e}$ is a unit vector of polarization (we assume polarization is linear). The equation for the wave function $\psi$ in a final state has a form

$$\left(\frac{\nabla^2}{2} - U(\mathbf{r}) + E\right)\psi = W\psi_i. \tag{12}$$

Here $\psi_i$ is the wave function of an initial electron state with the energy $E_i$ in a negative ion, $E$ is an electron energy in a final state after a photodetachment, $E = E_i + \omega$, $\omega$ is a frequency of the absorbed photon, $U(\mathbf{r})$ is a static field potential at the point $\mathbf{r}$.

For a homogeneous electric field $\mathbf{F}$ directed along the OZ axis, the stationary Green function $G_E$ is the solution to the equation

$$\left(\frac{\nabla^2}{2} + Fz + E\right)G_E(\mathbf{r},\mathbf{r}_1) = \delta(z-z_1)\delta(x-x_1)\delta(y-y_1), \tag{13}$$

where $\delta$ is the Dirac delta-function. The exact expression for the three-dimensional Green function is [37]

$$G_E(\mathbf{r},\mathbf{r}_1) = \frac{1}{2}\frac{1}{|\mathbf{r}-\mathbf{r}_1|}[\text{Ci}(\alpha_+)\text{Ai}'_i(\alpha_-) - \text{Ci}'_i(\alpha_+)\text{Ai}_i(\alpha_-)], \tag{14}$$

$$\alpha_\pm = -\left(\frac{1}{4F^2}\right)^{1/3}[2E + \mathbf{F}(\mathbf{r}-\mathbf{r}_1) \pm F|\mathbf{r}-\mathbf{r}_1|],$$

where $\text{Ai}(\alpha)$ and $\text{Ci}(\alpha)$ are the Airy functions, $\text{Ai}'(\alpha)$ and $\text{Ci}'(\alpha)$ are their derivatives [38]. The current at a large distance $r$ has the form

$$j \approx j_0 \text{Ai}^2\left[\frac{1}{\lambda_0}\left(\frac{\rho^2}{4z} - a\right)\right], \tag{15}$$

$\lambda_0 = (2F)^{-1/3}$, $a = E/F$, $z$ is the distance between negative ion and the center of screen, $\rho$ is the transverse coordinate. At the caustic surface ($\rho = \rho_{max} = \sqrt{4zE/F}$) the magnitude of the electronic current remains finite. In the classically accessible area

$$j \approx \frac{\sin^2\left[\frac{2}{3}\left(\frac{a}{\lambda_0}\right)^{3/2}\left(1-\rho^2/\rho_{max}^2\right)^{3/2} + \pi/4\right]}{\sqrt{1-\rho^2/\rho_{max}^2}}. \tag{16}$$

This expression is agreed well with the semiclassical result given by Eq. (9). It should be stressed that correct semiclassical formula (16) can be derived from Eq. (10) only with proper Maslov indexes. This fact sometimes was ignored even in serious publications [39].

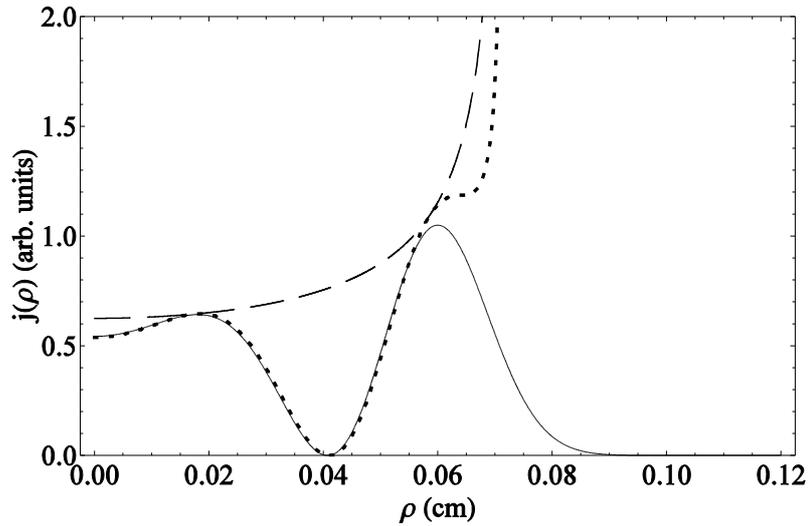

FIG. 1. Transverse distribution of a photocurrent in a homogeneous field: a dashed curve is a classical approximation; dot curve is a semiclassical result; a continuous curve is a quantum theory calculation.

In Fig. 1 the distribution of a photocurrent $j(\rho)$ is plotted for $F = 100$ V/m, $z = 0.5$ m, $E = 0.2 \text{ cm}^{-1}$ ($1 \text{ cm}^{-1} = 4.5 \cdot 10^{-6}$ a.u. $= 1.239 \cdot 10^{-4}$ eV). It is apparent that semiclassical approximation reproduces an exact solution far from the caustic surface. The classical result indicates the position of a caustic but interference pattern with rings of intensity is entirely absent. This result points clearly that classical motion calculations [40] have a very restricted applicability even in relatively simple cases.

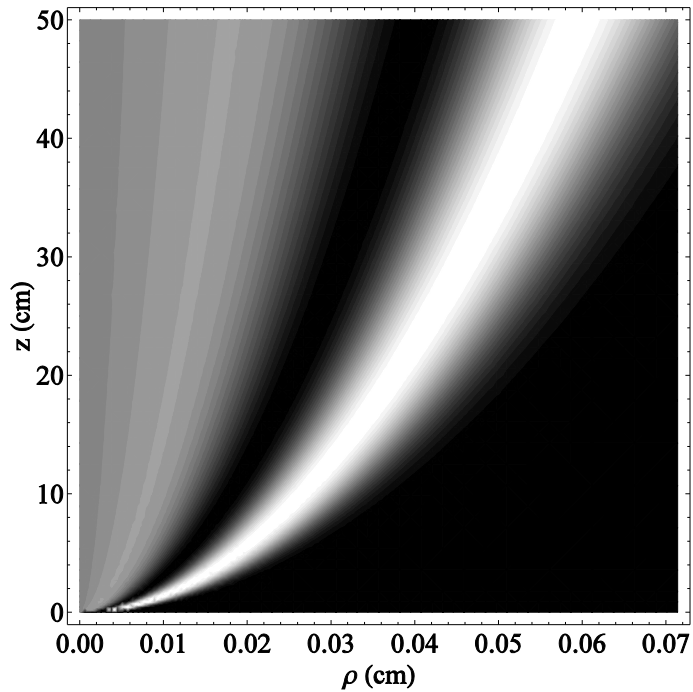

FIG. 2. Variation of distribution of a photocurrent with a distance to the source.

In Fig. 2 the change of a photocurrent distribution along the transverse direction $\rho$ with the growth of the distance $z$ is shown as a shadow graphic. A transversal extension of electron flux is demonstrated during its propagation along the electric field. These calculations and basic formulas give us data for comparison with alternative geometry. It is significant that complete analysis of a uniform field scheme is very useful for calculation of the distribution close to caustic even in more complicated cases [20, 41].

Investigation of other electric field configurations for more effective extension of the interference pattern at a smaller distance may be of interest. One of the most interesting cases is an electron photodetachment in the external repulsive Coulomb field $U(r) = \alpha/r$ as shown in Fig.3.

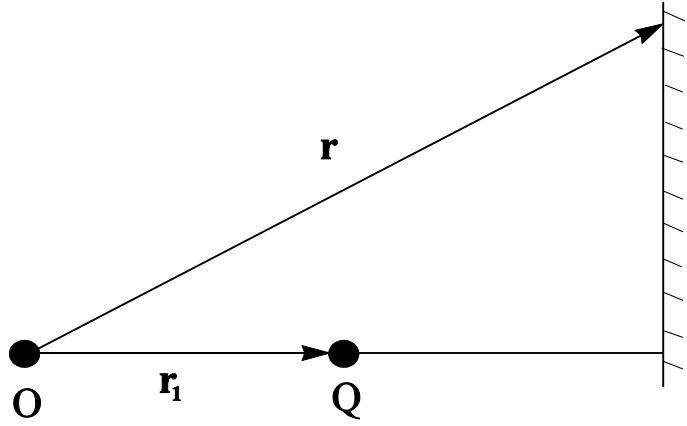

FIG. 3. The mutual position of the repulsive Coulomb center O and a negative ion Q.

Analytical representation for the Green function in the Coulomb field with a negatively charged point center is [42]

$$G_E(\mathbf{r},\mathbf{r}_1) = \frac{2\Gamma(1-i\eta)}{4\pi|\mathbf{r}-\mathbf{r}_1|} \times \frac{1}{ik}\left(-\frac{\partial}{\partial y}+\frac{\partial}{\partial x}\right)\cdot W_{i\eta,\frac{1}{2}}(-ikx) M_{i\eta,\frac{1}{2}}(-iky) =$$

$$= -\frac{\Gamma(1-i\eta)}{2\pi|\mathbf{r}-\mathbf{r}_1|}\begin{vmatrix} W_{i\eta,\frac{1}{2}}(-ikx) & M_{i\eta,\frac{1}{2}}(-iky) \\ W'_{i\eta,\frac{1}{2}}(-ikx) & M'_{i\eta,\frac{1}{2}}(-iky) \end{vmatrix}. \quad (17)$$

Here $\eta = -m\alpha/k$, $k = \sqrt{2mE}$, $x = r + r_1 + |\mathbf{r}-\mathbf{r}_1|$, $y = r + r_1 - |\mathbf{r}-\mathbf{r}_1|$, $W_{i\eta,\frac{1}{2}}$ and $M_{i\eta,\frac{1}{2}}$ are the Whittaker functions, $\Gamma(v)$ is gamma-function [43], vector $\mathbf{r}_1$ connects a repulsive center and a negative ion.

At a large distance $r$ from the Coulomb center, the radial density of a current $j_r$ has the asymptotic form

$$j_r = \frac{1}{m}\text{Im}\left[\psi*\frac{\partial \psi}{\partial r}\right] \approx \frac{k}{m}\text{Im}[\psi*\psi] = \frac{k}{m}|AG(\mathbf{r},\mathbf{r}_1)|^2 \sim$$

$$\sim \frac{1}{r^2}\left|M'_{i\eta,\frac{1}{2}}(-iky) - M_{i\eta,\frac{1}{2}}(-iky)\right|^2. \quad (18)$$

For a larger $r$ the term with derivative can be neglected. The multiplier $1/r^2$ is nearly constant at a large distance from the Coulomb center. It is instructive to consider the distribution in a transverse direction $\rho$, where

$$j(\rho) \sim \left|M_{i\eta,\frac{1}{2}}(-iky)\right|^2. \tag{19}$$

From the structure of the Green function (see Eq. (17)), it is clear that one-dimensional equation for the Coulomb potential

$$\frac{d^2\chi}{dr^2} + \left(k^2 - \frac{2m\alpha}{r}\right)\chi = 0 \tag{20}$$

is principal for accurate development of three-dimensional Green function in the repulsive Coulomb field. For more detailed analyses of Eq. (20) we replace variable: $v = r/\beta$. The new equation is of the form

$$\frac{d^2\chi}{dv^2} + \left(1 - \frac{Z}{v}\right)\chi = 0, \tag{21}$$

where $\beta = 1/k$, $Z = 2m\alpha$.

The semiclassical solution of the one-dimensional Schrödinger equation in classically inaccessible area ($v < Z$) is [44]

$$\chi(v) = \frac{A}{\sqrt{|p|}} \exp\left(-\left|\int_{r_0}^{r} p\, dr\right|\right). \tag{22}$$

In classically accessible area ($v > Z$) we have

$$\chi(v) = \frac{B}{\sqrt{p}} \cos\left(\int_{r_0}^{r} p\, dr - \pi/4\right). \tag{23}$$

The integral is calculated over the classical area, starting from a turning point $r_0 = 2m\alpha/k^2$. The phase shift $\pi/4$ arises as a result of reflection at the classical turning point. Calculating a phase, where $p(r) = \sqrt{k^2 - 2m\alpha/r}$, we have

$$S(r) = \int_{r_0}^{r} p(r)dr = kr_0\left(\sqrt{s(s-1)} - \ln\left(\sqrt{s-1} + \sqrt{s}\right)\right), \tag{24}$$

and $s = v/Z = r/r_0$. For the classically inaccessible region $p(r) = \sqrt{Z/v - 1}$, and

$$S(r) = \int_{r_0}^{r} p(r)dr = kr_0\left(s\sqrt{1/s - 1} + \arcsin(2s-1) - \pi/2\right). \tag{25}$$

Then the solution for this region ($v < Z$)

$$\chi = \frac{C}{(1/s-1)^{1/4}} \exp\left(-kr_0\left(s\sqrt{1/s-1} - \arcsin(2s-1) + \pi/2\right)\right), \qquad (26)$$

and in classically accessible area ($v > Z$)

$$\chi = \frac{C}{(1-1/s)^{1/4}} \sin\left(kr_0\left(\sqrt{s(s-1)} - \ln\left(\sqrt{s-1} + \sqrt{s}\right)\right) + \pi/4\right). \qquad (25)$$

The results of calculations, according to the semiclassical approximation, in comparison with the exact solution are shown in Fig. 4.

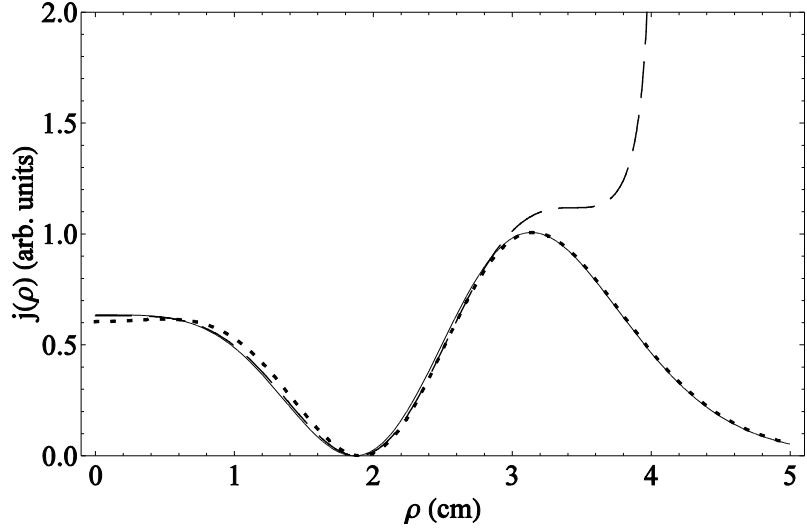

FIG. 4. Density of an electronic current for the following parameters: $z = 5$ cm, $|\mathbf{r}_1| = 2\times 10^{-4}$ cm, $E = 0.2$ cm$^{-1}$ and the electric field strength near negative ion $F = 100$ V/m. Semiclassical solution is shown by a dashed, uniform field approximation is dotted line, the exact solution is a continuous line.

Evidently, semiclassical approximation is good enough for classically accessible area. In the vicinity of the caustic, the semiclassical approximation tends to infinity, whereas the exact solution remains finite. Near caustic Eq. (21) has a form

$$\frac{d^2\chi}{dv^2} + \left(\frac{v-Z}{Z}\right)\chi = 0. \qquad (26)$$

This equation is identical to the case of uniform field with the solution

$$\chi(v) = \mathrm{Ai}\left[Z^{2/3}(v/Z - 1)\right]. \qquad (27)$$

A uniform field approximation gives a rather good result for all transversal distances and in combination with semiclassical approximation for $\rho < 2$ cm reproduces exact result with high accuracy. The approximation is so successful because the field in near screen area is not far from uniform. This special situation may be used for calculation of photocurrent in asymptotic region for more complicated combination of fields with cylindrical symmetry, when the caustic surfaces off the symmetry axis represent the most basic catastrophe, the fold. Electron waves undergoing reflection at the fold-type caustic surfaces locally resemble Airy functions [45] suggests an approximation scheme that works uniformly for all distances $\rho$. This is convenient for practical calculations of quantum mechanical interference because the main difficulty is shifted to the determination of caustic surface that is much more simple classical mechanics problem.

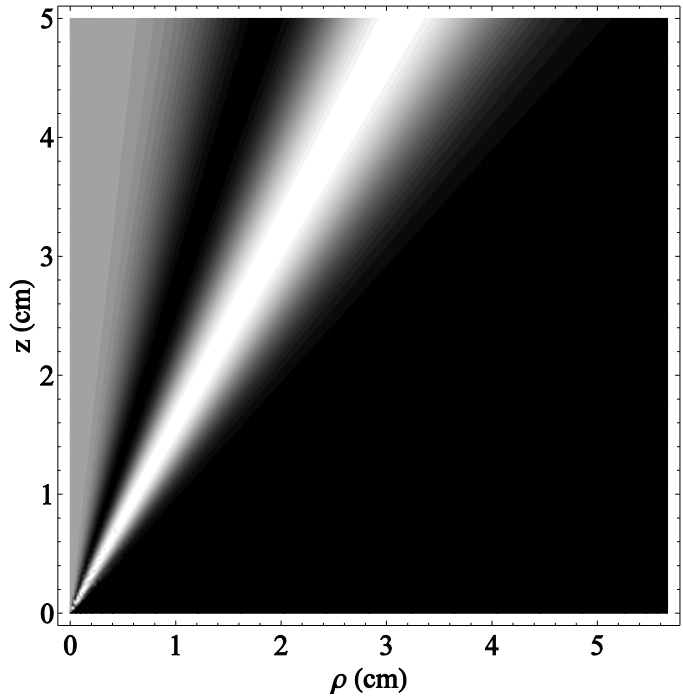

FIG. 5. Variation of a photocurrent $j(\rho)$ with the growth of $z$ as a function of transverse coordinate $\rho$.

Space distribution of an electron current is demonstrated in Fig. 5. The repulsive Coulomb field plays the role of projector and magnifies the initial distribution dramatically.

### 4. CONCLUDING REMARCS

Our calculations demonstrate new opportunities for a control of electron waves by a static electric field. Operating with the repulsive Coulomb field, it is possible to produce a multiply magnified projection of electron flux. For the one-electron wave packets this problem was discussed in [46]. A number of experimental and theoretical investigations were previously concentrated on the case of homogeneous electric field. The interference spot has the size of near micrometer under real experimental conditions. The trajectories are stretched along electric field direction, so that in the experiments the distance between the negative ion and registration plate is approximately one meter. However, electron wave interference has a universal physical nature and occurs in different field combinations. The newest experimental efforts are concentrated in two directions: modification of laser sources to two-color and short pulses and variation of field including magnetic field effects.

Photodetachment microscopy was carried out on a beam of iodine negative ions with a nanosecond pulse laser [47]. The photoelectron interferograms are recorded by means of a digital camera that images the light spots produced by the photoelectrons on a phosphor screen. Due to their sensitivity to the photoelectron energy, the recorded electron interferograms can be quantitatively analyzed to produce a measure of the electron affinity of iodine with an accuracy improved by more than a factor of 2 with respect to the best previous measurement.

A two-color laser technique was developed for photodetachment microscopy, by means of microwave modulation of a single-mode laser [48]. A phase modulation regime is achieved through an electro-optical crystal excited at the frequency 1.95 GHz. With the resulting two-color radiation, the photodetachment microscopy technique is applied to a beam of $S^-$ ions. It is shown that the superposition of the two resulting interference patterns can be used as a 'spectral vernier' to remove the uncertainty on the electric field and absolute energy scale. A measure of the electric field and of the electron affinity of sulfur is obtained.

The effect of an external static magnetic field of arbitrary orientation with respect to the electron affinity on the electron interference ring patterns, observed by the photodetachment microscope, has been studied both experimentally and theoretically [49]. The essential result of this work is invariance of the extreme interference phase whatever the direction and magnitude of the applied magnetic field, up to values 100 times larger than in the previous experimental study. This property can be applied to revise former electron affinity measurements.

The drawback to all these experiments the size of the imagine spot at the detector is small. An electrostatic lens that magnifies the images of an existing velocity map imaging apparatus up to a factor of 20 has been designed and implemented [50]. The lens can be used to vary the magnification while keeping the field strength in the interaction region constant. For the region of interest where magnification is required, for low energy electrons in a high external field, the lens does not add any observable aberrations to the imaging. The lens has been tested using slow photoelectrons resulting from the photoionization of metastable xenon atoms. The electrons are created in the interaction region at the crossing point of the laser and the molecular beam of metastable xenon atoms. However, the wave analysis of this lens that includes interference is not in hand up to now.

We propose to employ the repulsive Coulomb field, because it is effective projection scheme, for the purposes of an electron coherent microscopy. The total analytical solution of the problem is based on the Green function for the Coulomb field and expressed in terms of the Whittaker functions. New asymptotic solutions, with the help of global semiclassical approximation and simple uniform field approach, allow immediately calculate a photocurrent and can be used for more complicated configurations of static electric field with cylindrical symmetry. The suggested experimental geometry, based on nano size electrode technology [52], guarantees an effective magnification of the interference pattern. Inside the central part of the interference area, the pattern may be reproduced by means of much more simple semiclassical calculations. We assume the best configuration for a negative ion photodetachment microscope is two-field scheme with a combination of the homogeneous electric field and the repulsive Coulomb field.

---

[1] K. Heinz, U. Starke, and J. Bernhardt, Progress in Surface Science **64**, 163 (2000).
[2] M. R. A. Shegelski, M. Reid, and L. Pow, Ultramicroscopy **84**, 159 (2000).
[3] C. Blondel, C. Delsart, F. Dulieu, Phys. Rev. Lett. **77**, 37 (1996).
[4] C. Blondel, C. Delsart, F. Dulieu, and C. Valli, Eur. Phys. J. D **5**, 207 (1999).
[5] P. A. Golovinski, Optics and Spectr. **84**, 723 (1998).
[6] P. A. Golovinski, JETP **85**, 857 (1997).
[7] I. I. Fabrikant, Sov. Phys. JETP **52**, 1045 (1980).
[8] Y. N. Demkov, V. D. Kondratovich, and V. N. Ostrovskii, JETP Lett. **34**, 403 (1981).
[9] P. A. Golovinski, Phys. Lett. A **183**, 89 (1993).
[10] C. Blondel, C. Delsart, C. Valli, S. Yiou, M.R. Godefroid, and S. van Eck, Phys. Rev. A **64**, 052504 (2001).
[11] C. Valli, C. Blondel, C. Delsart, Phys. Rev. A **59**, 3809 (1999).
[12] C. Blondel, S. Berge, C. Delsart, Am. J. Phys. **69**, 810 (2001).
[13] C. Blondel, C. Delsart, and F. Goldfarb, J. Phys. B **34**, L281 (2001).
[14] C. Delsart, F. Goldfarb, C. Blondel, Phys. Rev. Lett. **89**, 183002 (2002).
[15] C. Valli, C. Blondel, and C. Delsart, Phys. Rev. A **59**, 3809 (1999).
[16] C. Blondel, C. Delsart, and F. Goldfarb, J. Phys. B **34**, L281 (2001).
[17] C. Blondel, W. Chaibi, C. Delsart, C. Drag, F. Goldfarb, and S. Krö ger, Eur. Phys. J. D

**33**, 335 (2005).
[18] W. Chaibi, C. Delsart, C. Drag, and C. Blondel, J Mol. Spectrosc. **239**, 11 (2006).
[19] C. Blondel, W. Chaibi, C. Delsart, and C. Drag, J. of Mod. Opt. **53**, 2605-2607 (2006)
[20] L. B. Zhao and J. B. Delos, Phys. Rev. A **81**, 53417 (2010).
[21] L. B. Zhao and J. B. Delos, Phys. Rev. A **81**, 53418 (2010).
[22] C. Nicole, H. L. Offerhaus, V. J. J. Vrakking, F. Lépine, and Ch. Dordas, Phys. Rev. Lett. **88**, 133001 (2002).
[23] P. A. Golovinski and A. A. Drobyshev, Proc. SPIE **7993**, 799311 (2010).
[24] M. L. Du and J. B. Delos, Phys. Rev. A **38,** 5609 (1988)
[25] H.-Y. Wong, A. R. P. Rau, and C. Y. Green, Phys. Rev. A **37**, 2393 (1988).
[26] Bo Gao, and A. F. Starace, Phys. Rev. A **42**, 5580 (1990).
[27] N. Y. Du, N. Y. Fabrikant, and A. F. Starace, Phys. Rev. A **48**, 2968 (1993).
[28] M.-Q. Bao, I. I. Fabrikant, and A. F. Starace, Phys. Rev. A **58,** 411 (1998).
[29] N. D. Gibson, M. D. Gasda, K. A. Moore, D. A. Zawistowski, and C. W. Walter, Phys. Rev. A **64**, 061403 (2001).
[30] I.I. Fabrikant, J. Phys. B **27**, 4545 (1994).
[31] I. I. Fabrikant, Sov. Phys. JETP **52**, 1045 (1980).
[32] I. I. Fabrikant, J. Phys. B **23**, 1139 (1990).
[33] V. L. Du, Phys. Rev. A **40**, 4983 (1989).
[34] V. I. Arnold, Mathematical Methods of Classical Mechanics (Springer-Verlag, New York, 2010).
[35] V. P. Maslov and M. V. Fedoryuk, Semiclassical Approximation in Quantum Mechanics (Reidel, Boston, 1981).
[36] V. B. Berestetskii, L. P. Pitaevskii, and E. M. Lifshitz, Quantum Electrodynamics (Butterworth-Heinemann, Oxford, 1982).
[37] B. Gottlieb, M. Kleber, and J. Krause, Z. Phys. A **339,** 201 (1991).
[38] M. Abramowitz and I. A. Stegun, Handbook of Mathematical Functions (National Bureau of Standards, Washington, 1972).
[39] Ch. Bordas, F. Lepine, C. Nicole, and V. J. J. Vrakking, Phys. Rev. A **68**, 012709 (2003).
[40] Ch. Bordas, Phys. Rev. A **58**, 400 (1998).
[41] Ch. Bracher, Kramer N., and J. B. Delos, Phys. Rev. A **73**, 062114 (2006).
[42] A. Baz', Ya. Zeldovich, and A. Perelomov, Scattering, Reactions and Decays in Nonrelativistic Quantum Mechanics (Israel Program for Scientific Translations, Jerusalem, 1969).
[43] E. T. Whittaker and G. N. Watson, A Course of Modern Analysis (Cambridge University Press, NY, 2002).
[44] L.D. Landau and E.M. Lifshitz, Quantum Mechanics: Non-Relativistic Theory (Butterworth-Hei nemann, Oxford, 1981).
[45] T. Poston and I. Stewart, Catastrophe Theory and its Applications, Pitman, London, 1978.
[46] V. P. Bykov and V. O. Turin, Laser Phys. **8**, 1039 (1998).
[47] R.J. Pelaez, C. Blondel, C. Delsart, and C. Drag, J. Phys. B: At. Mol. Opt. Phys. **42**, 125001 (2009).
[48] C. Drag, W. Chaibi, C. Delsart, C. Blondel, Opt. Comm. **275**, 190 (2007).
[49] W. Chaibi, R.J. Pelaez, C. Blondel, C. Drag, and C. Delsart, Eur. Phys. J. D **58**, 29 (2010).
[50] H. L. Offerhaus, C. Nicole, F. Lépine, C. Bordas, F. Rosca-Pruna, and M. J. J. Vrakking, Review of Scientific Instruments **72**, 3245 (2001).
[51] A. V. Eletskii, Phys. Usp. **53**, 863 (2010).